\documentclass[conference]{IEEEtran}

\usepackage{cite}
\usepackage{amsmath,amssymb,amsfonts}
\usepackage{graphicx}
\usepackage{booktabs}
\usepackage{textcomp}
\usepackage{xcolor}
\usepackage[nocheckfootnote]{flushend} 

\newcommand{\rev}[1]{{\color{black}#1}} 
\newif\ifincludeappendix
\includeappendixtrue
\ifdefined\submissiononly\includeappendixfalse\fi

\begin{document}

\title{Collision-Resilient All-to-All Broadcast Scheduling in Underwater Acoustic Networks}

\author{
Guoliang Weng, Enqi Zhang, Lizhao You, Deqing Wang, and Liqun Fu
}

\author{\IEEEauthorblockN{
		  Guoliang Weng$^\diamond$,
          Enqi Zhang$^\dagger$,
          Lizhao You$^{\dagger}$,
          Deqing Wang$^\dagger$, and Liqun Fu$^\dagger$}
	\IEEEauthorblockA{
        $^\diamond$School of Artifical Intelligence, Xiamen Unviersity, China\\
        $^\dagger$School of Informatics, Xiamen University, China}
\IEEEcompsocitemizethanks{}
}

\maketitle

\begin{abstract}
This paper studies all-to-all broadcast scheduling for underwater swarms with collision-resilient physical layers that can decode asynchronously overlapping packets. 
Propagation-aware collision-free schedulers coordinate packet arrivals but do not account for this reception capability.
We first design a data-driven neural reception model that captures per-interferer power and timing effects. The model improves prediction accuracy over SINR baselines while admitting an exact mixed-integer linear programming (MILP) formulation.
The resulting neural-network-embedded MILP (NN-MILP) optimizes transmission times while explicitly encoding schedule-dependent overlaps and the predictor's inference, enabling globally optimal scheduling for small-scale instances.
For larger networks, a two-stage ant colony optimization (ACO) heuristic combines low-cost transmission-order search with neural-aware timing refinement.
Simulation results show that ACO shortens broadcast rounds compared with existing schedulers. It approaches the NN-MILP optimum on small-scale instances while supporting larger networks at substantially lower computational cost.
\end{abstract}


\section{Introduction}

Cooperative multi-agent systems (e.g., multi-AUV swarms) rely on frequent information exchange to support localization, formation control, distributed sensing, and collective decision-making~\cite{yang2021survey,cai2023cooperative}. They communicate over long-range acoustic links, so one information-exchange round takes the form of a single-hop all-to-all (A2A) broadcast, in which every node transmits one state-update packet intended for all other nodes. The scheduled duration of such a round directly bounds how often network-wide state information can be refreshed, and thus the responsiveness of swarm coordination.

Efficient A2A broadcast is challenging in underwater acoustic networks
because of long propagation delays and half-duplex
operation~\cite{stojanovic2009underwater}. 
Propagation-aware schedulers exploit differences in packet arrival times
across receivers by optimizing transmission order and
timing~\cite{park2010parallel,qiu2023optimal,liu2023adaptive,he2025spatial}.
However, their collision-free reception constraints exclude overlapping
packets, limiting the concurrency available to the scheduler.

Collision-resilient underwater PHYs offer an opportunity to relax this
restriction. Asynchronous multiuser detection in direct-sequence
CDMA~\cite{stojanovic2006cdma} and the Zadoff--Chu-sequence-based waveform
ZCMod~\cite{zhang2025zcmod} demonstrate that overlapping underwater
transmissions can be decoded. Such capabilities make some overlaps usable
rather than necessarily destructive, potentially shortening A2A broadcast
rounds beyond collision-free scheduling. Exploiting this opportunity requires
a scheduler that identifies which receiver-specific overlaps satisfy a
prescribed reception-reliability requirement and jointly optimizes
transmission timing.

This cross-layer opportunity raises three coupled link-layer challenges.
First, \emph{the scheduler needs an accurate and efficient reception model for asynchronously overlapping packets}. Decoding depends jointly on received powers and receiver-specific arrival offsets, whose effects are not adequately captured by conventional SINR-based abstractions, as illustrated in Section~\ref{sec:predictor}. Prior underwater MAC studies model
concurrent reception using near--far capture relationships or statistical
aggregate-interference descriptions~\cite{diamant2017nearfar,guan2021stochastic},
but these abstractions do not preserve the deterministic, per-interferer
arrival-time structure induced by a candidate schedule.
We adopt a data-driven approach to learn packet reception success probability from decoding outcomes obtainable through simulations or experiments. 
A shared encoder and permutation-invariant pooling preserve each interferer's power and timing information while accommodating unordered interference sets. The resulting predictor enables reliability-aware evaluation of candidate schedules.

Second, \emph{globally optimal scheduling is challenging because of coupled timing constraints and is further complicated by neural reception prediction}. 
Changing one transmission time alters overlap relationships at multiple receivers while affecting half-duplex feasibility. Incorporating the predictor additionally requires its schedule-dependent inputs and neural inference to be represented within the optimization constraints.
We address these coupled difficulties by choosing a reception-model architecture that balances prediction accuracy with exact mixed-integer linear representability~\cite{fischetti2018deep}.
The resulting neural-network-embedded mixed-integer linear programming (NN-MILP) formulation jointly encodes transmission timing, receiver-specific interference contexts, and predicted reception probabilities. 
It minimizes broadcast completion time subject to half-duplex and predicted-reliability constraints. For small-scale instances, the solver can find and certify a globally optimal schedule under the given model and fixed predictor.

Third, \emph{efficient scheduling becomes increasingly difficult as the network grows}. The NN-MILP requires rapidly increasing numbers of variables and constraints, making exact optimization impractical for larger networks. 
A heuristic must also control its computational cost: evaluating many candidate schedules requires repeated reception predictions across multiple receivers.
We therefore propose a two-stage ant colony optimization (ACO) heuristic~\cite{dorigo1997antcolony}. In the first stage, ants search over transmission orders using a low-cost, propagation-aware collision-free decoder. This decoder provides a surrogate score without repeatedly invoking the neural reception model.
Since an order that performs well under collision-free scheduling may not remain best when overlaps are allowed, we retain a set of promising orders with diverse structures rather than only the surrogate-best order.
In the second stage, a neural-aware decoder refines transmission times for these selected orders, allowing overlaps that satisfy the predicted-reliability constraints.
Each resulting schedule is checked against all half-duplex and predicted-reliability constraints, and the valid schedule with the shortest broadcast completion time is returned.
This separation limits expensive neural-aware evaluations while preserving broad exploration of transmission orders.

We implement a simulator that combines a network scheduler with a physical-layer simulator and evaluate the scheduling methods on randomly deployed 3D underwater networks.
Simulation results show that ACO reduces mean broadcast completion time by $23.5\%$ and $19.5\%$ compared with random-order and greedy scheduling, respectively, while remaining within $4.5\%$ of the NN-MILP optimum on tractable instances. It also achieves $29.7\%$--$39.9\%$ shorter broadcast rounds than the collision-free optimum at the common tested node counts, highlighting the benefit of allowing reliable overlapping receptions. Although ACO requires more computation than random and greedy scheduling, it avoids the steep runtime growth of exact optimization and offers a favorable tradeoff between scheduling quality and computational cost.

The main contributions are summarized as follows:
\begin{enumerate}
    \item A permutation-invariant reception model that predicts packet
    success probability under receiver-specific asynchronous interference
    and supports MILP formulation.

    \item An NN-MILP formulation that minimizes broadcast completion
    time under half-duplex and predicted-reliability constraints.

    \item A two-stage ACO scheduler that combines low-cost order search
    with neural-aware refinement, achieving near-benchmark performance
    on small instances and scaling to larger networks.
\end{enumerate}

\section{System Model}
\label{sec:system}

\rev{This section defines the network models and formulates
the collision-resilient A2A broadcast scheduling problem.}

\subsection{Network Model}
We consider a static single-hop underwater acoustic network (UAN) comprising
\(N\) half-duplex nodes, indexed by
\(\mathcal{N}=\{0,\ldots,N-1\}\), that share a common acoustic channel and
are deployed at known three-dimensional locations
\(\mathbf{q}_i\in\mathbb{R}^{3}\). During each scheduling frame, every node
broadcasts one state-update packet of duration \(L\) directly to all the other
\(N-1\) nodes; hence, each ordered pair \((i,j)\), \(i\neq j\), represents an
intended reception. We consider one periodic information-refresh
frame and do not include acknowledgments, relaying, or retransmissions. 

The scheduler assigns each node \(i\) a nonnegative continuous
transmission start time \(s_i\). Let
\(d_{ij}=\lVert\mathbf{q}_i-\mathbf{q}_j\rVert_2\) and
\(\tau_{ij}=d_{ij}/c\) denote the distance and one-way propagation delay from
node \(i\) to node \(j\), respectively, where \(c\) is the acoustic propagation
speed. 
Accordingly, node \(j\) receives the packet broadcast by node \(i\) over \([s_i+\tau_{ij},\,s_i+\tau_{ij}+L)\), so the same broadcast generally reaches different receivers at different times. Let $\tau_{max}$ denote the maximum propagation delay across all node pairs. 
Let \(P^{\mathrm{rx}}_{ij}\) denote the estimated power of node \(i\)'s
packet at receiver \(j\). It can be obtained from link measurements at the receiver, and is assumed to be same during one scheduling frame. 

Each node operates in strict half-duplex mode and therefore cannot transmit
while receiving another node's packet. Let
\(\mathcal{T}_j=[s_j,\,s_j+L)\) denote the transmission interval of node \(j\),
and let
\(\mathcal{R}_{ij}=[s_i+\tau_{ij},\,s_i+\tau_{ij}+L)\) denote the interval over
which node \(j\) receives the packet from node \(i\). A schedule is
half-duplex feasible only if
\begin{equation}
    \mathcal{T}_j \cap \mathcal{R}_{ij}=\varnothing,
    \qquad \forall i,j\in\mathcal{N},\ i\neq j.
    \label{eq:half_duplex}
\end{equation}
This receiver-side constraint does not prohibit concurrent transmissions in
general: multiple nodes may transmit simultaneously provided that no node's
transmission interval overlaps any reception required at that node.

\noindent{\bf Receiver-Specific Asynchronous Interference.}
To characterize receiver-side interference, consider a desired reception
\(i\to j\) under the schedule
\(\mathbf{s}=(s_i)_{i\in\mathcal{N}}\). Any node
\(k\in\mathcal{N}\setminus\{i,j\}\) acts as an interfering sender whenever
its packet reception interval at node \(j\) overlaps \(\mathcal{R}_{ij}\).
Accordingly, we define the overlapping-interferer set as
\begin{equation}
    \mathcal{O}_{ij}(\mathbf{s})=
    \left\{k\in\mathcal{N}\setminus\{i,j\}:\
    \mathcal{R}_{kj}\cap\mathcal{R}_{ij}\neq\varnothing\right\}.
    \label{eq:overlap_set}
\end{equation}
For each \(k\in\mathcal{O}_{ij}(\mathbf{s})\), its arrival-time offset relative
to the desired packet is
\(\Delta_{k\mid ij}=(s_k+\tau_{kj})-(s_i+\tau_{ij})\).
Since acoustic
propagation delays depend on the receiver location, the same pair of
broadcasts may overlap at one receiver but not at another; hence, the
interference context must be constructed separately for every ordered
sender--receiver pair.

The complete asynchronous interference context of reception \(i\to j\)
can be represented as
\begin{equation}
    \mathcal{I}_{ij}(\mathbf{s})=
    \left(P^{\mathrm{rx}}_{ij},
    \left\{(\Delta_{k\mid ij},P^{\mathrm{rx}}_{kj}):
    k\in\mathcal{O}_{ij}(\mathbf{s})\right\}\right).
    \label{eq:interference_context}
\end{equation}
Let \(f_L\) denote the reception behavior of the underlying
collision-resilient PHY for packet duration \(L\). The corresponding packet
success probability is abstractly written as
\begin{equation}
    p^{\mathrm{PHY}}_{ij}(\mathbf{s})=
    f_L\!\left(\mathcal{I}_{ij}(\mathbf{s})\right).
    \label{eq:phy_reliability}
\end{equation}

\subsection{Problem Formulation}
Since the activity of node \(i\) ends only after its packet reaches its farthest receiver, the one-round broadcast
scheduling time is defined as
\begin{equation}
    C_{\max}(\mathbf{s})=
    \max_{i\in\mathcal{N}}\Big\{s_i+L+
    \underbrace{\max_{j\in\mathcal{N}\setminus\{i\}}\tau_{ij}}_{h_i}\Big\},
    \label{eq:broadcast_round_time}
\end{equation}
where \(h_i\) is the propagation tail of node \(i\).
The collision-resilient all-to-all broadcast scheduling problem is then
formulated as
\begin{equation}
\begin{aligned}
    \min_{\mathbf{s}\in\mathbb{R}_{+}^{N}}\quad
    & C_{\max}(\mathbf{s})\\
    \mathrm{s.t.}\quad
    & \mathcal{T}_j\cap\mathcal{R}_{ij}=\varnothing,
      && \forall i,j\in\mathcal{N},\ i\neq j,\\
    & p^{\mathrm{PHY}}_{ij}(\mathbf{s})\geq\rho,
      && \forall i,j\in\mathcal{N},\ i\neq j,
\end{aligned}
\label{eq:semantic_scheduling_problem}
\end{equation}
where the second requires every link to meet the target
reliability \(\rho\in(0,1)\). 

\rev{
Solving this problem is challenging because a single transmission-time decision affects receptions at multiple nodes. Changing one start time can alter both half-duplex feasibility and the membership and arrival offsets of overlapping interferers. The resulting reliability constraints
therefore couple continuous timing decisions with discrete interference patterns. 
Moreover, the PHY reception function $f_L$ does not generally admit a tractable analytical expression, and repeatedly evaluating it directly is too costly during scheduling.

These difficulties motivate a data-driven reception model that supports both efficient prediction and optimization-compatible representation.
Our key idea is to learn the reception behavior using a neural architecture that admits an exact mixed-integer linear representation.
We then jointly encode transmission timing, interference membership, predictor-input construction, and neural inference in a MILP formulation.
This converts the learned reception-probability requirements into solver-compatible constraints, allowing transmission times to be optimized without repeatedly evaluating the underlying PHY.

The resulting formulation provides an exact optimization method for the specified scheduling model with the fixed predictor.
However, prediction errors may cause its feasible schedules and optimal solution to differ from those of the original PHY-based problem. 
Thus, optimality under the learned model does not guarantee global optimality or reception reliability under the true PHY.
Nevertheless, the learned model enables principled scheduling
decisions that account for asynchronous interference beyond
conventional collision-free abstractions.
}

\section{Data-Driven Reception Prediction Model}
\label{sec:predictor}




\rev{
This section introduces SE-SetNet, a data-driven surrogate of the PHY reception function $f_L$. Given a candidate schedule $\mathbf{s}$, SE-SetNet predicts the success probability of each of the $N(N-1)$ intended receptions. 
For each sender--receiver pair $i\to j$, it uses the desired received power and the powers and relative arrival offsets of all overlapping interferers to estimate $p^{\mathrm{PHY}}_{ij}(\mathbf{s})$. 
These predictions provide an approximation of the reception probabilities used in \eqref{eq:semantic_scheduling_problem}.

Although data-driven methods have been used to predict underwater
channel quality, bit-error rate, and communication
performance~\cite{jing2022adaptive,seol2025deep},
these studies cannot handle reception predictions as schedule-dependent reliability constraints, which requires a model that is both accurate and compatible with scheduling optimization.
} 

\subsection{Model Design}
\rev{
The model must satisfy two requirements: accurate prediction under asynchronous interference and compatibility with exact MILP embedding.
Each reception involves an unordered set of overlapping packets whose received powers and arrival offsets jointly affect decoding.
We therefore encode interferers individually and aggregate their representations through permutation-invariant pooling.
The encoder and prediction head use affine layers and ReLU activations, while the output probability is obtained through a sigmoid function. This architecture allows the predicted-reliability constraint to be represented exactly in the scheduling MILP.
}

To reduce the surrogate complexity, we admit only schedules in which every reception has at most \(K\) overlapping interferers, where more than $K$ overlaps are treated as an error. 
Under this limit, we construct the predictor SE-SetNet as follows.
Let
\(q_{ij}(\mathbf{s})=|\mathcal{O}_{ij}(\mathbf{s})|\) and
\(\mathcal{K}_{ij}=\mathcal{N}\setminus\{i,j\}\).
\rev{
For each overlapping interferer $k$, we construct a normalized
feature vector containing the desired received power, relative
arrival offset, and interfering received power:
\begin{equation}
\mathbf{v}_{ijk}=
\begin{bmatrix}
(P^{\mathrm{rx}}_{ij}-\mu_P)/\sigma_P\\
(\Delta_{k\mid ij}-\mu_t)/\sigma_t\\
(P^{\mathrm{rx}}_{kj}-\mu_I)/\sigma_I
\end{bmatrix}.
\label{eq:set_tokens}
\end{equation}}
Here \(\mu_P,\mu_t,\mu_I\) and \(\sigma_P,\sigma_t,\sigma_I>0\) are the
means and standard deviations of the three features over the training set. 
\rev{The normalization parameters are computed from the training data and held fixed during scheduling. Since received powers are also fixed, only the arrival-offset feature changes with transmission times. Its normalization requires only division by a fixed constant and subtraction of a fixed offset, so the feature construction in \eqref{eq:set_tokens} can be represented by linear equality
constraints without approximation.}

A shared neural network encoder \(\phi_{\boldsymbol\theta}\)
with ReLU activations maps each token to an embedding.
The pooled representation and head input are
\begin{equation}
\begin{aligned}
\mathbf{r}_{ij}
&=\frac{1}{K}
\sum_{k\in\mathcal{O}_{ij}(\mathbf{s})}
\phi_{\boldsymbol\theta}(\mathbf{v}_{ijk}),\\
\mathbf{x}_{ij}(\mathbf{s})
&=[\kappa_{ij},\mathbf{r}_{ij}^{\mathsf T},
q_{ij}(\mathbf{s})/K]^{\mathsf T}.
\end{aligned}
\label{eq:nn_input_vector}
\end{equation}
Sum pooling makes the representation invariant to
interferer ordering. The fixed divisor \(K\) and explicit
count preserve information about interference-set size.
\rev{The prediction head also receives the desired-link power $\kappa_{ij}=(P^{\mathrm{rx}}_{ij}-\mu_P)/\sigma_P$ and the normalized interferer count $q_{ij}/K$. These inputs explicitly describe the target link and interference-set size. When no interferer overlaps the desired packet, the pooled vector is zero, but the head still predicts reception success from the desired-link power.}

A NN head with a ReLU hidden layer and an affine
output layer produces the logit and predicted reception probability:
\begin{equation}
z_{ij}=\psi_{\boldsymbol\theta}(\mathbf{x}_{ij}(\mathbf{s})),
\qquad
\widehat p_{ij}(\mathbf{s})=(1+\exp(-z_{ij}))^{-1}.
\label{eq:reliability_model}
\end{equation}


\subsection{Model Training and Selection}
\rev{
Training data consist of reception contexts and their decoding
outcomes, obtained through simulation or experiments. For each
context, repeated trials yield an empirical success probability, defined as the fraction of packets decoded without bit errors.  
After training, the network weights and normalization parameters are
frozen. The frozen SE-SetNet is then used to evaluate schedule-dependent
reception probabilities in the predictor-constrained scheduling problem
and its NN-MILP and ACO solvers.
}

We compare SE-SetNet with the conventional block-wise SINR model used in NS3, the simplest interfer-counting-based SINR model (Any-Overlap), a simple MLP-based NN model (Flat-MLP), and an attention-based NN model (SetAttn). \rev{In our implementation, contexts are generated from the NS-3 network simulator and evaluated using the MATLAB PHY simulator.}
Table~\ref{tab:model_comparison} compares prediction error and model size. SE-SetNet has lower MAE and RMSE than the analytical baselines and Flat-MLP, and uses fewer parameters than Flat-MLP. SetAttn achieves the lowest errors, but its nonlinear attention operations complicate exact MILP embedding. We use SE-SetNet because its affine layers, ReLU activations, and fixed pooling admit a compact MILP representation, and its relatively high performance.




\begin{table}[t]
    \centering
    \caption{Prediction accuracy and model size.}
    \label{tab:model_comparison}
    \footnotesize
    \setlength{\tabcolsep}{4pt}
    \begin{tabular}{lrrr}
        \toprule
        Model & Parameters & MAE $\downarrow$ & RMSE $\downarrow$\\
        \midrule
        Block-wise SINR & -- & 0.148966 & 0.268919\\
        \shortstack[l]{Any-Overlap}
            & -- & 0.161963 & 0.289872\\
        Flat-MLP & 1,249 & 0.039890 & 0.074159\\
        \textbf{SE-SetNet}
            & \textbf{197}
            & \textbf{0.028269}
            & \textbf{0.048020}\\
        SetAttn & 431,489 & 0.023052 & 0.040006\\
        \bottomrule
    \end{tabular}
    \vspace{-0.2in}
\end{table}

\rev{
\subsection{Predictor-Constrained Scheduling}
Replacing the PHY reception probabilities in
\eqref{eq:semantic_scheduling_problem} with SE-SetNet
yields the following surrogate scheduling problem:
\begin{equation}
\begin{aligned}
\min_{\mathbf{s}\in\mathbb{R}_{+}^{N}}\quad
& C_{\max}(\mathbf{s})\\
\mathrm{s.t.}\quad
& \mathcal{T}_j\cap\mathcal{R}_{ij}=\varnothing,
  && \forall i\neq j,\\
& q_{ij}(\mathbf{s})\leq K,
  && \forall i\neq j,\\
& \widehat p_{ij}(\mathbf{s})\geq\rho,
  && \forall i\neq j.
\end{aligned}
\label{eq:optimization_problem}
\end{equation}
The predicted probabilities vary with the transmission times through the resulting interference contexts. 
Section~\ref{sec:milp} presents an NN-MILP formulation that explicitly encodes these dependencies.
}

\section{\rev{NN-MILP Formulation}}
\label{sec:milp}


\rev{
We convert the predictor-constrained scheduling problem into an NN-MILP.
The MLPs' affine layers and fixed pooling are represented by linear equalities, while ReLU activations and interference masking use mixed-integer linear constraints~\cite{fischetti2018deep}.
The sigmoid reliability requirement becomes an equivalent logit threshold. 
These representations are combined with transmission-timing and interference-selection constraints, as defined below.
}

\subsection{Timing and Interference Selection}

Let \(H\) be a finite scheduling horizon and impose
\(0\leq s_i\leq H\) for every node. For each reception
\(i\to j\), a binary \(u_{ij}\) selects whether node \(j\)
transmits before or after receiving the desired packet:
\begin{equation}
\begin{aligned}
C_{\max}&\geq s_i+\tau_{ij}+L,\\
\rev{u_{ij}=1}&\rev{\Rightarrow s_j+L\leq s_i+\tau_{ij}},\\
\rev{u_{ij}=0}&\rev{\Rightarrow s_i+\tau_{ij}+L\leq s_j}.
\end{aligned}
\label{eq:milp_timing_core}
\end{equation}
\rev{The first inequality bounds the end of each reception, while the
indicator constraints prevent simultaneous transmission and reception
at a node. The intervals may meet at their endpoints without overlapping.}

\rev{The overlapping-interferer set is not known before scheduling.
We therefore represent all \(N-2\) candidates in
\(\mathcal K_{ij}=\mathcal N\setminus\{i,j\}\) and use binary states
to determine their temporal relationships with the desired packet.}
For every candidate \(k\in\mathcal{K}_{ij}\), the relative
arrival offset is
\(\Delta_{k\mid ij}=s_k-s_i+\tau_{kj}-\tau_{ij}\).
Binary variables \(b_{ijk},o_{ijk},a_{ijk}\) select the before,
selected-overlap, and after states, respectively:
\begin{equation}
\begin{gathered}
b_{ijk}+o_{ijk}+a_{ijk}=1,\\
q^{\mathrm{sel}}_{ij}=\sum_{k\in\mathcal{K}_{ij}}o_{ijk}\leq K,\\
b_{ijk}=1\Rightarrow\Delta_{k\mid ij}\leq-L,\\
o_{ijk}=1\Rightarrow-L\leq\Delta_{k\mid ij}\leq L,\\
a_{ijk}=1\Rightarrow\Delta_{k\mid ij}\geq L.
\end{gathered}
\label{eq:milp_selection_core}
\end{equation}
\rev{The selected count \(q^{\mathrm{sel}}_{ij}\) enforces the predictor's
interferer limit. At \(\Delta_{k\mid ij}=\pm L\), either adjacent state
is admissible. The MILP evaluates the selected interferer set, which
equals the strict-overlap set away from these endpoint contacts.}

\rev{To construct bounded encoder inputs for all candidates, introduce
\(e_{ijk}\in[-L,L]\) with indicator constraints
\(o_{ijk}=1\Rightarrow e_{ijk}=\Delta_{k\mid ij}\) and
\(o_{ijk}=0\Rightarrow e_{ijk}=0\).
Replacing the offset in \eqref{eq:set_tokens} with \(e_{ijk}\) gives
\(\widetilde{\mathbf v}_{ijk}\), using the same fixed normalization.
Received powers remain constant; only the timing feature changes
through fixed scaling and shifting.}

\subsection{Exact Encoder Masking and Set Pooling}

\rev{Zeroing an inactive candidate's offset does not remove its encoder
output: the remaining power features and network biases can produce
nonzero embeddings. We therefore mask each output by its selection
variable before pooling.}

Each candidate is evaluated by the same frozen encoder,
\(\mathbf{h}_{ijk}=\phi_{\boldsymbol\theta}
(\widetilde{\mathbf{v}}_{ijk})\).
Because inputs differ, each evaluation has separate
intermediate variables and constraints, while sharing all
encoder weights. The final encoder layer uses ReLU, giving
\(0\leq h_{ijk,r}\leq U_{ijk,r}\), where \(r=1,\ldots,4\)
and \(U_{ijk,r}\) is a valid upper bound over the scheduling
domain. The product \(m_{ijk,r}=o_{ijk}h_{ijk,r}\) is
represented exactly by
\begin{equation}
\begin{aligned}
0\leq m_{ijk,r}&\leq h_{ijk,r},\\
m_{ijk,r}&\leq U_{ijk,r}o_{ijk},\\
m_{ijk,r}&\geq h_{ijk,r}-U_{ijk,r}(1-o_{ijk}).
\end{aligned}
\label{eq:milp_mask_core}
\end{equation}
When \(o_{ijk}=1\), these inequalities retain the encoder
output; otherwise they force the masked contribution to zero.

The masked embeddings form the pooled representation and
head input:
\begin{equation}
\begin{aligned}
r_{ij,r}&=\frac{1}{K}
\sum_{k\in\mathcal{K}_{ij}}m_{ijk,r},\\
\mathbf{x}^{\mathrm{head}}_{ij}
&=[\kappa_{ij},\mathbf{r}_{ij}^{\mathsf T},
q^{\mathrm{sel}}_{ij}/K]^{\mathsf T}.
\end{aligned}
\label{eq:milp_pool}
\end{equation}
\rev{Fixed sum pooling preserves permutation invariance and remains
linear in the masked embeddings. Together with the normalized desired-link
power and selected count, it forms the head's six-dimensional input.}

\subsection{ReLU and Reliability Embedding}

For any encoder or hidden-head neuron \(\nu\), let
\(\eta_{\nu}=\mathbf{w}_{\nu}\mathbf{x}_{\nu}+b_{\nu}\)
be its affine preactivation, with valid bounds
\(\underline\eta_{\nu},\overline\eta_{\nu}\).
For \(\underline\eta_{\nu}<0<\overline\eta_{\nu}\), a binary
\(\beta_{\nu}\) yields the exact ReLU representation
\begin{equation}
\begin{aligned}
y_{\nu}&\geq0,\qquad y_{\nu}\geq\eta_{\nu},\\
y_{\nu}&\leq\overline\eta_{\nu}\beta_{\nu},\\
y_{\nu}&\leq\eta_{\nu}-\underline\eta_{\nu}(1-\beta_{\nu}).
\end{aligned}
\label{eq:milp_relu_core}
\end{equation}
When \(\beta_{\nu}=0\), the inequalities enforce
\(y_{\nu}=0\) and \(\eta_{\nu}\leq0\); when
\(\beta_{\nu}=1\), they enforce \(y_{\nu}=\eta_{\nu}\)
and \(\eta_{\nu}\geq0\). Thus, each binary chooses an
activation branch while preserving the neuron's continuous
output within that branch.
If \(\overline\eta_{\nu}\leq0\), the neuron is fixed at
\(y_{\nu}=0\); if \(\underline\eta_{\nu}\geq0\), set
\(y_{\nu}=\eta_{\nu}\). These stable cases require no
activation binary. Bounds are obtained by interval propagation
from the scheduling domain, including \(e_{ijk}\in[-L,L]\),
rather than empirical training extrema, so the encoding covers
all admissible network inputs.

Applying these constraints to both encoder layers and the
head's hidden layer leaves only affine operations in the
output logit \(z_{ij}\). For \(0<\rho<1\), sigmoid monotonicity
enforces the required reception probability through
\begin{equation}
z_{ij}\geq\gamma_{\rho},
\qquad
\gamma_{\rho}=\log\frac{\rho}{1-\rho},
\qquad \forall i\neq j.
\label{eq:nn_logit_threshold}
\end{equation}
\rev{This threshold, together with masking and ReLU constraints, encodes
the selected-set reliability test exactly, including empty selected sets.
The complete formulation is provided in Appendix~A.}

\section{Ant-Colony-Based Heuristic Scheduling}
\label{sec:aco}



\rev{
As the network grows, the NN-MILP becomes expensive because it jointly
represents transmission timing, receiver-specific interference states,
and neural-network inference. We therefore develop a two-stage ant
colony optimization (ACO) heuristic. The first stage searches node
orders using a low-cost, propagation-aware collision-free surrogate.
The second stage selects diverse promising orders and refines their
transmission times using the reception predictor. Each resulting
schedule is fully validated, and the valid schedule with the smallest
broadcast completion time is returned.
}

\subsection{Surrogate-Guided Order Search}

Each ant constructs a permutation
\(\boldsymbol{\pi}=(\pi_1,\ldots,\pi_N)\).
A deterministic decoder sequentially places nodes to produce
a collision-free, half-duplex-feasible schedule. Its completion
time \(\widetilde C(\boldsymbol{\pi})\) accounts for propagation
delays and serves as the surrogate score. \rev{This score ranks orders
without neural evaluation; it neither certifies predicted reliability
nor restricts the refined schedules to collision-free operation.}

Order construction combines edge pheromone, which captures
preferred node adjacencies, and position pheromone, which
captures preferred node positions. A propagation-aware
heuristic guides the selection of candidate \(i\) at
position \(t\):
\begin{equation}
\eta_{ti}=
\frac{1+\lambda h_i/S}{1+\Delta C_{ti}/S},
\label{eq:aco_heuristic}
\end{equation}
where \(h_i\) is the propagation tail,
\(\Delta C_{ti}\) is the increase in surrogate completion
time, and \(S=L+\max_{p,q}\tau_{pq}\).
The weight \(\lambda\) controls the preference for long
propagation tails. \rev{Ants sample the next node with probability
proportional to the combined pheromone and heuristic scores.}
Pheromones are updated through evaporation
and reinforcement of the iteration-best and global-best orders.

\subsection{\rev{Elite Selection and Neural-Aware Timing Refinement}}

Because surrogate rankings may change when reliable overlaps
are allowed, the second stage evaluates a structurally diverse
elite set. Two orders based on mean propagation delays and
propagation tails initialize the set. \rev{Additional elites are selected
from high-ranking archived orders using a score combining surrogate rank
with max--min diversity in directed adjacencies and node positions.}

\rev{Each elite permutation specifies a sequential insertion order,
not necessarily the chronological order of the resulting transmissions.} Candidate
start times span zero to the collision-free anchor and
include coarse samples and samples around propagation-shifted
activity boundaries. These event-centered candidates explore
placements where half-duplex or interference relationships
can change without densely scanning the time interval.

Each candidate is checked for half-duplex feasibility,
interferer-count limits, and predicted reliability.
\rev{All affected receptions are reevaluated, including previously
scheduled receptions disturbed by the new transmission.} Among feasible candidates, the decoder
selects the one minimizing partial broadcast completion time.
The collision-free anchor must pass the same checks;
an elite is discarded if no feasible insertion exists.

Each completed schedule is validated by reconstructing all
receiver contexts and checking \rev{half-duplex feasibility,
interferer-count limits, and predicted reliability}. The validated schedule
with the smallest \(C_{\max}\) is returned. If all elite
decodes fail, the best surrogate schedule is considered
only after passing the same full validation. \rev{If the fallback also
fails, the algorithm reports that no feasible schedule was found.}
\rev{The algorithm details are provided in Appendix~B.}

\section{Performance Evaluation}
\label{sec:evaluation}

\rev{This section describes the simulation setup and evaluates broadcast
completion time and computational cost under different network sizes
and packet durations.}

\begin{figure*}[t!]
    \centering
    \begin{minipage}[t]{\dimexpr(\textwidth-2\columnsep)/3\relax}
        \vspace{0pt}
        \centering
        \parbox[b][0.8\linewidth][b]{\linewidth}{\centering
            \includegraphics[width=\linewidth]{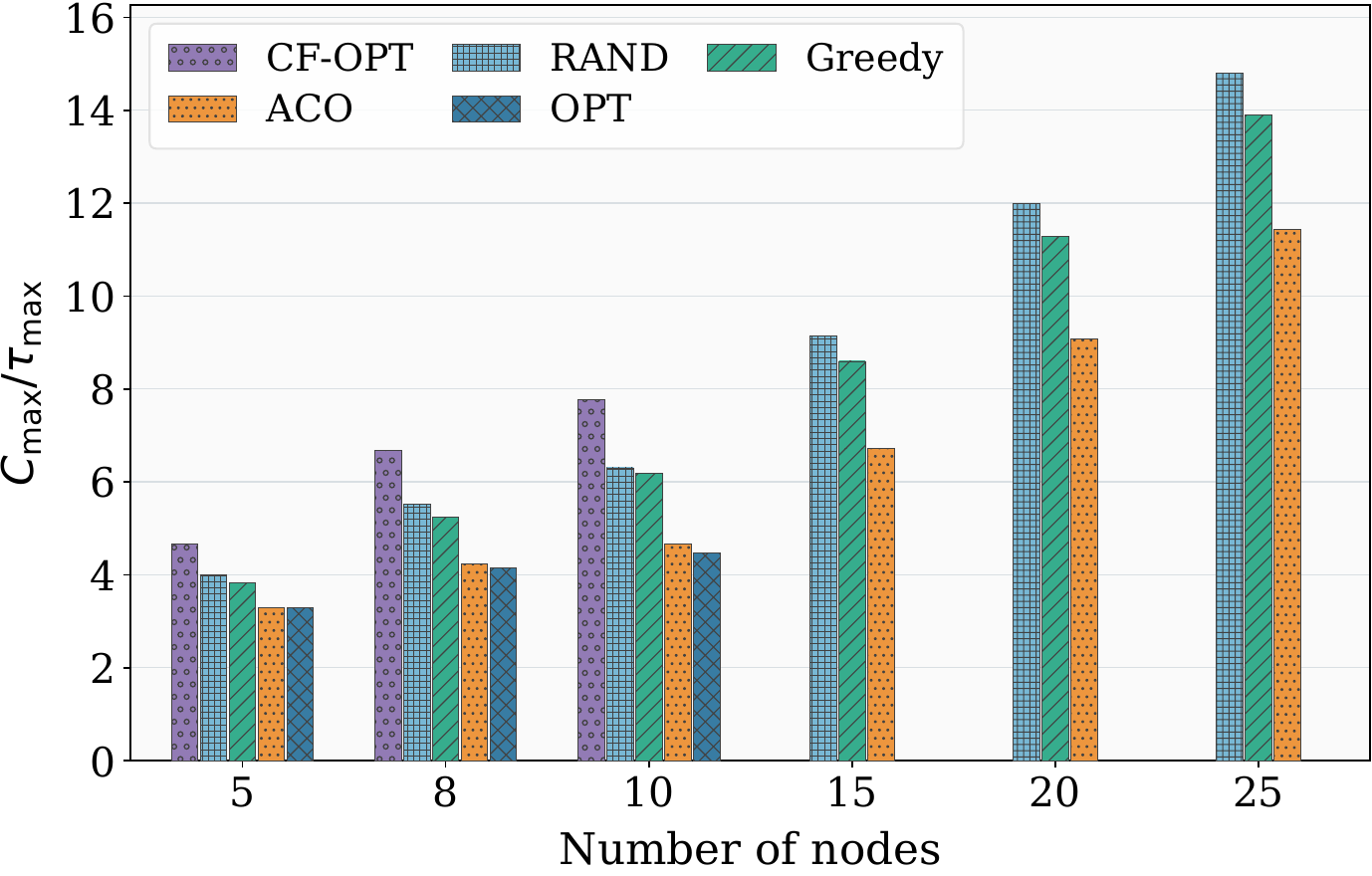}}
        \vspace{1mm}
        \begin{minipage}[t][2.8em][t]{\linewidth}
            \caption{Normalized broadcast completion time versus network size for five scheduling methods.}
            \label{fig:cmax_time}
        \end{minipage}
    \end{minipage}\hfill
    \begin{minipage}[t]{\dimexpr(\textwidth-2\columnsep)/3\relax}
        \vspace{0pt}
        \centering
        \parbox[b][0.8\linewidth][b]{\linewidth}{\centering
            \includegraphics[width=\linewidth]{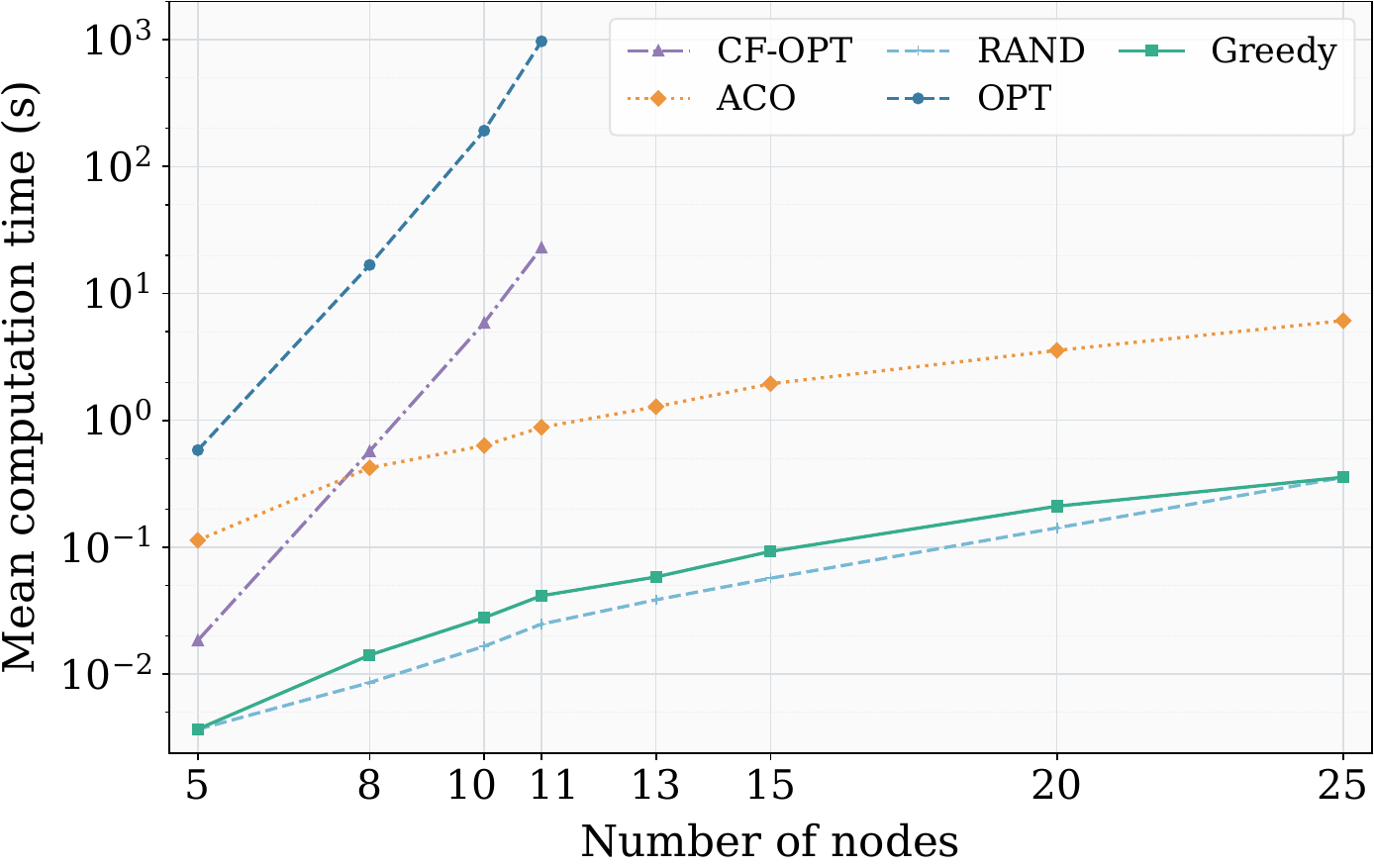}}
        \vspace{1mm}
        \begin{minipage}[t][2.8em][t]{\linewidth}
            \caption{Mean computation time versus network size for five scheduling methods.}
            \label{fig:comp_time}
        \end{minipage}
    \end{minipage}\hfill
    \begin{minipage}[t]{\dimexpr(\textwidth-2\columnsep)/3\relax}
        \vspace{0pt}
        \centering
        \parbox[b][0.8\linewidth][b]{\linewidth}{\centering
            \includegraphics[width=\linewidth]{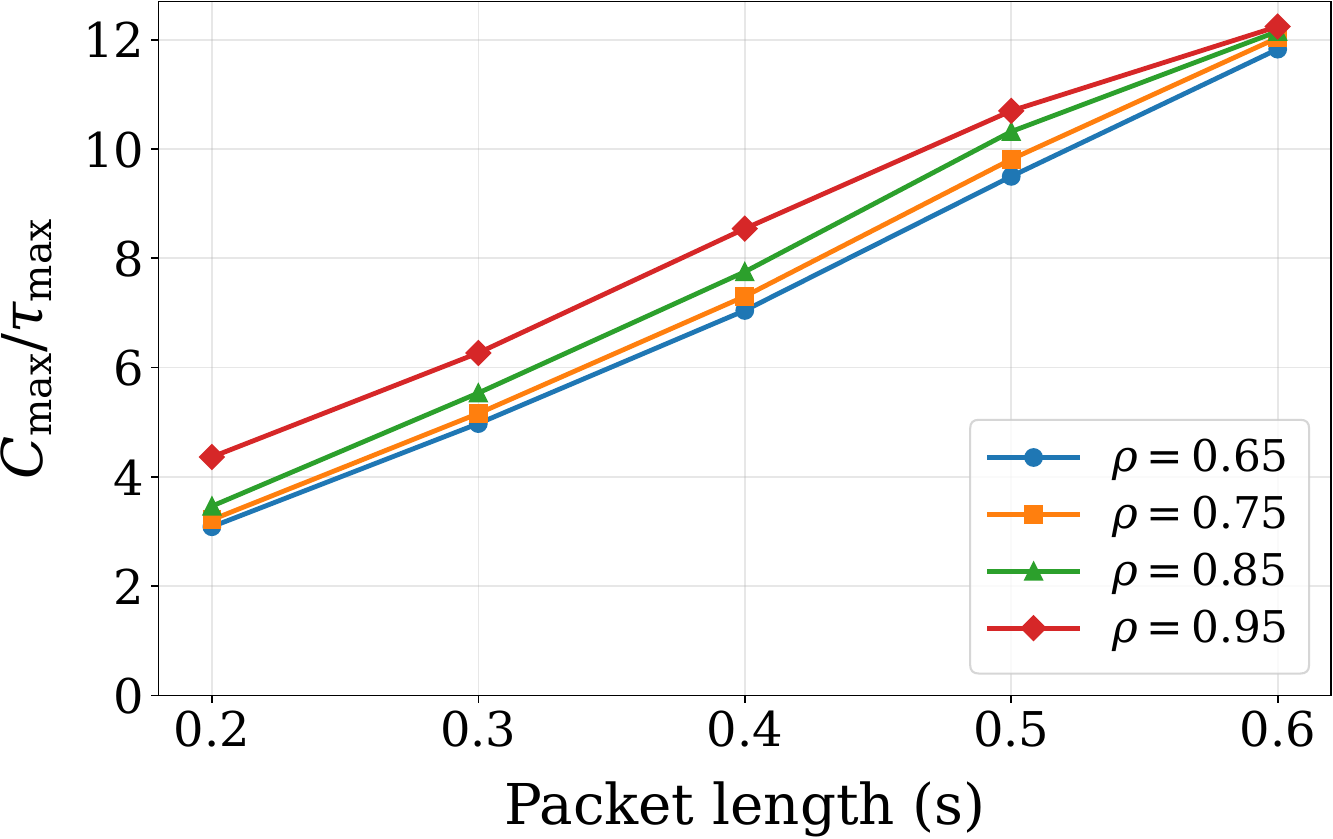}}
        \vspace{1mm}
        \begin{minipage}[t][2.8em][t]{\linewidth}
            \caption{Normalized completion time versus packet duration
            for different reliability thresholds.}
            \label{fig:packet_time}
        \end{minipage}
    \end{minipage}
    \vspace{-0.1in}
\end{figure*}

\subsection{Simulation Setup}
Nodes are randomly deployed in a
$1000\,\mathrm{m}\times1000\,\mathrm{m}\times500\,\mathrm{m}$
underwater region.
The acoustic propagation speed is $1500\,\mathrm{m/s}$.
Large-scale attenuation combines spherical spreading, referenced
to $1\,\mathrm{m}$, with Thorp absorption evaluated at a carrier
frequency of $20\,\mathrm{kHz}$~\cite{thorp1967analytical}.
The received power of each directed link is assumed to be estimable
at the receiver and available to the scheduler. In the simulations,
these powers are computed using the same propagation model and
waveform normalization as in predictor training and remain fixed
within each scheduling frame.

The PHY simulator is configured for PN-based direct-sequence CDMA
with a spreading factor of 127, 4-DPSK modulation, and rate-$1/2$
LDPC coding. The carrier frequency and sampling rate are
$20\,\mathrm{kHz}$ and $60\,\mathrm{kHz}$, respectively.
Interfering waveforms are independently generated, attenuated
according to their link distances, and shifted by their relative
arrival offsets before being superposed with the desired signal.
AWGN is added once to the real-valued passband waveform after
superposition and before downconversion. With transmit waveform
power normalized to unity, the noise variance is
$1.08\times10^{-6}$ in normalized units.
For overlapping receptions in the interference-limited regime,
decoding performance depends primarily on relative received powers
and arrival offsets, rather than a common scaling of all transmit
powers. Unless otherwise stated, the scheduling packet duration and the
predicted per-reception reliability threshold are
$L=0.30\,\mathrm{s}$ and $\rho=0.95$, respectively.

We compare the following scheduling methods. RAND and GREEDY
construct schedules by inserting one transmitter at a time.
For each candidate node, the earliest feasible transmission time is determined from the transmissions already scheduled.
Overlapping receptions are allowed.

\begin{enumerate}
    \item \textit{Random-sequence scheduling (RAND):}
    Nodes are selected uniformly at random from the unscheduled
    nodes and inserted at their earliest feasible transmission times.

    \item \textit{Greedy scheduling (GREEDY):}
    We adopt the greedy strategy used in~\cite{he2025spatial} and adapt the insertion feasibility check to our reception model. At each step, every unscheduled node is tentatively inserted
    at its earliest feasible transmission time. The node yielding the smallest $C_{\max}$ for the updated partial
    schedule is selected.

    \item \textit{Proposed NN-MILP (OPT):}
    Gurobi solves the reliability-aware scheduling formulation defined in Section~\ref{sec:milp}.
    Instances with certified optimal solutions provide the
    optimal reference.

    \item \textit{Proposed ACO heuristic (ACO):}
    The two-stage scheduler defined in Section~\ref{sec:aco} combines low-cost transmission-order
    search with neural-aware timing refinement of diverse
    promising orders.

    \item \textit{Collision-free optimal scheduling (CF-OPT):}
    Transmission times are optimized subject to collision-free
    reception. This benchmark
    quantifies the potential benefit of allowing reliable
    overlapping receptions.
\end{enumerate}

All scheduling algorithms are implemented in Python 3.11, and
the PHY simulation is implemented in MATLAB. Experiments are
conducted on a workstation with an AMD Ryzen7 7745HX processorand 16\,GB of RAM.
The NN-MILP is solved using Gurobi 13.0.2 with a time limit of
3600\,s per instance. A feasible solution returned at the time
limit is not treated as a certified optimum.

All methods use the same topology realizations, link parameters, and the same frozen reception predictor and reliability threshold. The predictor's training procedure is described in Section~\ref{sec:predictor}. For each configuration, results are averaged over 20 independent random topology realizations.





\subsection{Simulation Results}

\subsubsection{Comparison of Scheduling Methods}

Fig.~\ref{fig:cmax_time} compares normalized broadcast completion time,
$C_{\max}/\tau_{\max}$, as labeled in the plot. Completion time increases with
network size for every method over its displayed range. CF-OPT and OPT are shown only for $N=5$,
$8$, and $10$ due to computational limitations, whereas RAND, GREEDY, and ACO extend to $N=25$ due to their lower computational requirements.

ACO consistently outperforms RAND and GREEDY across all six
tested node counts, reducing normalized broadcast completion time
by $23.5\%$ and $19.5\%$ on average, respectively. Since these methods
use the same reception predictor and reliability threshold, the
improvements demonstrate the benefit of the proposed two-stage
search over random ordering and locally greedy insertion.
ACO also approaches the exact optimum on tractable instances. Its gaps to OPT are only $0.1\%$, $2.0\%$, and
$4.5\%$ for $N=5$, $8$, and $10$, respectively. These small gapsindicate that the heuristic retains most of the scheduling gain achievable under the NN-MILP model.

Finally, the comparison with CF-OPT highlights the benefit of collision-resilient scheduling. Despite being a heuristic, ACO achieves $29.7\%$--$39.9\%$ shorter broadcast rounds than the collision-free optimum at the three common node counts.
The result shows that optimizing within a collision-free feasible region can leave substantial temporal
reuse unexploited.

Fig.~\ref{fig:comp_time} reports mean computation time on a logarithmic
scale. OPT rises from $0.58\,\mathrm{s}$ at $N=5$ to $971\,\mathrm{s}$ at
$N=11$. CF-OPT also grows rapidly, from $0.018\,\mathrm{s}$ to
$23\,\mathrm{s}$ across the same range, and neither exact formulation is reported beyond $N=11$. ACO requires $0.11\,\mathrm{s}$ at $N=5$ and $0.88\,\mathrm{s}$ at $N=11$, which is approximately three orders of magnitude less than OPT at the same node count.
RAND and GREEDY are faster than ACO throughout the displayed range, reaching about $0.36\,\mathrm{s}$ at $N=25$, compared with $6.1\,\mathrm{s}$ for ACO. 
Thus, ACO incurs additional search cost but obtains shorter broadcast rounds, while avoiding the steep runtime growth of exact optimization on the tested instances. 

\subsubsection{Impact of Packet Duration}

Fig.~\ref{fig:packet_time} shows normalized completion time for packet
durations $L=0.2$--$0.6\,\mathrm{s}$ and thresholds
$\rho\in\{0.65,0.75,0.85,0.95\}$. All four curves increase monotonically
with $L$. The $\rho=0.65$ curve rises from about $3.1$ to
$11.8$, while the $\rho=0.95$ curve rises from about $4.4$ to $12.2$.
Longer packets extend transmission and reception intervals and leave less temporal flexibility for scheduling.


At every displayed packet duration, a stricter threshold yields a
longer broadcast round, consistent with fewer overlap configurations passing
the predicted-reliability test. However, the separation between thresholds
does not grow monotonically with packet duration. The gap between
$\rho=0.65$ and $0.95$ is about $1.3$--$1.5$ in normalized completion time
for $L=0.2$--$0.4\,\mathrm{s}$, but narrows to about $0.4$ at
$L=0.6\,\mathrm{s}$. Thus, the effect of the threshold on completion time is smaller at the longest packet duration in this experiment. 

\section{Conclusion}


We present a collision-resilient A2A broadcast scheduling framework
for underwater acoustic networks. It combines a neural reception
model, an NN-MILP formulation, and a two-stage ACO scheduler to
coordinate overlapping transmissions under half-duplex and
predicted-reliability constraints. Simulation results demonstrate
shorter broadcast rounds than the tested baselines, near-optimal
performance on small instances, and improved scalability over exact
optimization. Future work will develop distributed scheduling
algorithms to reduce reliance on centralized coordination.

\bibliographystyle{IEEEtran}
\bibliography{references}

\ifincludeappendix
\newpage
\appendices
\numberwithin{equation}{section}

\rev{

\section{Complete NN--MILP Formulation}
\label{app:complete_nnmilp}

This appendix details the NN-MILP in Section~\ref{sec:milp}, using the
notation of Sections~\ref{sec:system} and~\ref{sec:predictor}.
The following subsections describe timing bounds, interference selection,
neural-network embedding, and the consolidated optimization problem.

\subsection{Timing Bounds and Half-Duplex Constraints}
\label{app:milp_timing}
The key idea is to express half-duplex operation as a choice between
two temporal orders: a node either finishes transmitting before a reception starts or starts transmitting after that reception ends. Binary variables select the order, while finite timing bounds define the scheduling domain.


Let \(\tau_{\max}=\max_{i,j}\tau_{ij}\). We have
\begin{equation}
 H=(N-1)(L+\tau_{\max}),\qquad
 C^{\mathrm{ub}}=H+\tau_{\max}+L,
 \label{eq:milp_horizon}
\end{equation}
and impose
\begin{equation}
 \begin{aligned}
 0\leq s_i&\leq H, & &\forall i,\\
 L\leq C_{\max}&\leq C^{\mathrm{ub}},\\
 C_{\max}&\geq s_i+\tau_{ij}+L, & &\forall i\ne j.
 \end{aligned}
 \label{eq:milp_time_bounds}
\end{equation}
Binary variables \(f_i\) satisfy \(\sum_i f_i=1\) and
\(f_i=1\Rightarrow s_i=0\), ensuring that at least one node starts at
zero without fixing its identity or excluding simultaneous starts.
Shifting all start times by their minimum preserves relative arrival
offsets and half-duplex feasibility.

For every intended reception, a binary variable $u_{ij}$ selects whether node $j$ transmits before or after receiving node $i$'s packet:
\begin{equation}
 \begin{aligned}
 u_{ij}=1&\Rightarrow s_j+L\leq s_i+\tau_{ij},\\
 u_{ij}=0&\Rightarrow s_i+\tau_{ij}+L\leq s_j.
 \end{aligned}
 \label{eq:milp_half_duplex}
\end{equation}

\subsection{Interferer Selection and Effective Offsets}
\label{app:milp_selection}
Because the interferer set depends on transmission times, it cannot
be fixed before optimization. We represent all candidate interferers
and use binary states to determine their temporal relationships
with the desired packet. Effective offsets then provide bounded
encoder inputs for both selected and inactive candidates.

For a desired reception $i\to j$, let
$\mathcal{K}_{ij}=\mathcal{N}\setminus\{i,j\}$ denote the set of
all $N-2$ candidate interferers. For each $k\in\mathcal{K}_{ij}$,
introduce binary variables $b_{ijk}$, $o_{ijk}$, and $a_{ijk}$
for the before, selected-overlap, and after states, respectively. 
They satisfy
\begin{equation}
 \begin{gathered}
 \Delta_{k\mid ij}=s_k-s_i+\tau_{kj}-\tau_{ij},\\
 b_{ijk}+o_{ijk}+a_{ijk}=1,\qquad
 q^{\mathrm{sel}}_{ij}=\sum_{k\in\mathcal{K}_{ij}}o_{ijk}\leq K,\\
 b_{ijk}=1\Rightarrow\Delta_{k\mid ij}\leq-L,\\
 o_{ijk}=1\Rightarrow-L\leq\Delta_{k\mid ij}\leq L,\\
 a_{ijk}=1\Rightarrow\Delta_{k\mid ij}\geq L.
 \end{gathered}
 \label{eq:milp_interference_states}
\end{equation}
All $N-2$ potential interferers are considered. The selected set
$\mathcal{S}_{ij}=\{k:o_{ijk}=1\}$ contains
$q^{\mathrm{sel}}_{ij}\leq K$ interferers. 
It equals the strict-overlap set
\(\mathcal{O}_{ij}(\mathbf{s})\) except possibly at
\(\Delta_{k\mid ij}=\pm L\), where either adjacent state is feasible.

The encoder input uses an effective offset \(e_{ijk}\):
\begin{equation}
 \begin{gathered}
 -L\leq e_{ijk}\leq L,\\
 o_{ijk}=1\Rightarrow e_{ijk}=\Delta_{k\mid ij},\qquad
 o_{ijk}=0\Rightarrow e_{ijk}=0.
 \end{gathered}
 \label{eq:milp_effective_offset}
\end{equation}
Each candidate interferer uses the standardized encoder input
\begin{equation}
 \widetilde{\mathbf{v}}_{ijk}=
 \begin{bmatrix}
 (P^{\mathrm{rx}}_{ij}-\mu_P)/\sigma_P\\
 (e_{ijk}-\mu_t)/\sigma_t\\
 (P^{\mathrm{rx}}_{kj}-\mu_I)/\sigma_I
 \end{bmatrix}.
 \label{eq:milp_encoder_token}
\end{equation}
This replaces the offset in \eqref{eq:set_tokens} with \(e_{ijk}\), using
the same normalization. For inactive candidates, \(e_{ijk}=0\) is auxiliary;
masking below removes their encoded contributions. All candidates use
shared encoder weights but separate variables and constraints.

\subsection{Exact Encoder Masking and Set Pooling}
\label{app:milp_masking}
An inactive candidate can produce a nonzero encoder output because
of network biases, even when its effective offset is zero.
We therefore mask each encoded output using its selection variable
before pooling. This preserves selected contributions and removes
inactive ones exactly.

Let \(\mathbf{h}_{ijk}=\phi_{\boldsymbol\theta}
(\widetilde{\mathbf{v}}_{ijk})\). Both encoder layers use ReLU, so each
output component satisfies \(0\leq h_{ijk,r}\leq U_{ijk,r}\), where
\(r=1,\ldots,4\) and the upper bounds follow from interval propagation.
Introduce \(m_{ijk,r}=o_{ijk}h_{ijk,r}\) through
\begin{equation}
 \begin{aligned}
 0\leq m_{ijk,r}&\leq h_{ijk,r},\\
 m_{ijk,r}&\leq U_{ijk,r}o_{ijk},\\
 m_{ijk,r}&\geq h_{ijk,r}-U_{ijk,r}(1-o_{ijk}).
 \end{aligned}
 \label{eq:milp_mask}
\end{equation}
If \(o_{ijk}=1\), the true-offset encoding is retained; if \(o_{ijk}=0\),
the contribution is exactly zero, irrespective of encoder biases. Setting
the inactive raw offset to zero alone would not achieve this masking.
The head input is
\begin{equation}
 r_{ij,r}=\frac{1}{K}\sum_{k\in\mathcal{K}_{ij}}m_{ijk,r},\qquad
 \mathbf{x}^{\mathrm{head}}_{ij}=
 [\kappa_{ij},\mathbf{r}_{ij}^{\mathsf T},q^{\mathrm{sel}}_{ij}/K]^{\mathsf T}.
 \label{eq:app_milp_pool}
\end{equation}
When \(\mathcal{S}_{ij}=\mathcal{O}_{ij}(\mathbf{s})\), this equals the
predictor input in \eqref{eq:nn_input_vector}. At contacts, it follows
Subsection~\ref{app:milp_selection}'s selected-set convention and may differ
from the strict-overlap prediction. Both representations are
permutation-invariant, requiring no slot-assignment binaries.

\subsection{ReLU and Reliability Embedding}
\label{app:milp_relu}

Each ReLU has two linear branches, which can be selected by a
binary variable using valid bounds on its input.
For the final sigmoid output, we avoid approximating the function:
its monotonicity converts the probability requirement into an
equivalent linear threshold on the logit.

For either an encoder layer or the hidden head layer, let
\(\boldsymbol\eta=W\mathbf{x}+\mathbf{b}\). Given valid componentwise
bounds \(\underline{\mathbf{x}},\overline{\mathbf{x}}\), propagate
\begin{equation}
 \begin{aligned}
 \underline{\boldsymbol\eta}
 &=W^+\underline{\mathbf{x}}+W^-\overline{\mathbf{x}}+\mathbf b,\\
 \overline{\boldsymbol\eta}
 &=W^+\overline{\mathbf{x}}+W^-\underline{\mathbf{x}}+\mathbf b,
 \end{aligned}
 \label{eq:milp_interval_bounds}
\end{equation}
where \(W^+=\max(W,0)\) and \(W^-=\min(W,0)\) elementwise.
The bounds originate from the scheduling domain, including
\(e_{ijk}\in[-L,L]\), rather than empirical training extrema. The
implementation pads affine bounds outward to accommodate roundoff.

Index each encoder/head ReLU neuron by \(\nu\), including its network
copy and layer. For an unstable neuron with
\(\underline\eta_{\nu}<0<\overline\eta_{\nu}\), a binary activation
variable \(\beta_{\nu}\) gives the exact representation
\begin{equation}
 \begin{gathered}
 y_{\nu}\geq0,\qquad y_{\nu}\geq\eta_{\nu},\\
 y_{\nu}\leq\overline\eta_{\nu}\beta_{\nu},\qquad
 y_{\nu}\leq\eta_{\nu}-\underline\eta_{\nu}(1-\beta_{\nu}),\\
 \beta_{\nu}\in\{0,1\}.
 \end{gathered}
 \label{eq:relu_linearization}
\end{equation}
Stable neurons require no binaries: set \(y_\nu=0\) if
\(\overline\eta_\nu\leq0\), or \(y_\nu=\eta_\nu\) if
\(\underline\eta_\nu\geq0\). Apply these constraints to both encoder
layers and the hidden head layer. The affine output logit satisfies
\begin{equation}
 z_{ij}\geq\gamma_{\rho},\qquad
 \gamma_{\rho}=\log\frac{\rho}{1-\rho},\qquad\forall i\ne j.
 \label{eq:app_nn_logit_threshold}
\end{equation}
By sigmoid monotonicity, this threshold is exact and also applies to
empty selected sets.

\subsection{Consolidated Optimization Problem}

Let \(\Xi\) collect all timing, binary, and neural-inference variables.
Index encoder neurons by \(\nu=(\mathrm{enc},i,j,k,\ell,t)\), with
\(\ell\in\{1,2\}\) and \(t=1,\ldots,8\) or \(1,\ldots,4\), respectively,
and head hidden neurons by \(\nu=(\mathrm{head},i,j,t)\),
\(t=1,\ldots,16\). Let \(\mathcal V\) contain these ReLU indices,
excluding the affine output. The vectors \(\mathbf y^{e,\ell}_{ijk}\)
and \(\mathbf y^h_{ij}\) collect their outputs \(y_\nu\), with inputs
\(\mathbf x_\nu=\mathbf x^{e,\ell}_{ijk}\) and
\(\mathbf x_\nu=\mathbf x^h_{ij}\), respectively. Each neuron uses its
layer's fixed weight row \(\mathbf w_\nu\) and bias \(b_\nu\).
Encoder and head weights are shared across copies; neuron variables are
separate. Output-layer parameters are \(\mathbf w^o\in\mathbb R^{16}\)
and \(b^o\in\mathbb R\). Received powers and normalization parameters
also remain fixed during scheduling. Define
\begin{equation*}
\begin{aligned}
\mathcal{A}&=\{(i,j)\in\mathcal{N}^{2}:i\neq j\},\\
\mathcal{I}&=\{(i,j,k):(i,j)\in\mathcal{A},\ k\in\mathcal{K}_{ij}\},\\
\mathcal{V}_{u}&=\{\nu\in\mathcal{V}:
\underline{\eta}_{\nu}<0<\overline{\eta}_{\nu}\},\\
\mathcal{V}_{0}&=\{\nu\in\mathcal{V}:
\overline{\eta}_{\nu}\leq0\},\\
\mathcal{V}_{+}&=\{\nu\in\mathcal{V}:
\underline{\eta}_{\nu}\geq0\}.
\end{aligned}
\end{equation*}
The constraint groups \(\mathcal{C}_{\mathrm{hd}}\),
\(\mathcal{C}_{\mathrm{int}}\), and \(\mathcal{C}_{\mathrm{mask}}\)
are given by \eqref{eq:milp_half_duplex},
\eqref{eq:milp_interference_states}--\eqref{eq:milp_effective_offset},
and \eqref{eq:milp_mask}, respectively;
\(\widetilde{\mathbf{v}}_{ijk}\) is defined in
\eqref{eq:milp_encoder_token}. With elementwise
\(\mathbf{w}_{\nu}^{+}=\max(\mathbf{w}_{\nu},0)\) and
\(\mathbf{w}_{\nu}^{-}=\min(\mathbf{w}_{\nu},0)\), the fixed neuron
bounds are
\begin{equation}
\begin{aligned}
\underline{\eta}_{\nu}
&=\mathbf{w}_{\nu}^{+}\underline{\mathbf{x}}_{\nu}
+\mathbf{w}_{\nu}^{-}\overline{\mathbf{x}}_{\nu}+b_{\nu},\\
\overline{\eta}_{\nu}
&=\mathbf{w}_{\nu}^{+}\overline{\mathbf{x}}_{\nu}
+\mathbf{w}_{\nu}^{-}\underline{\mathbf{x}}_{\nu}+b_{\nu}.
\end{aligned}
\label{eq:app_neuron_bounds}
\end{equation}

Using these neuron bounds, we combine the timing, interference-selection,
and neural-network constraints into the complete NN-MILP below.
All indices refer to the sets defined above, and variables are continuous
unless explicitly declared binary.
\newpage
\begin{equation}
\begin{aligned}
\min_{\Xi}\quad
&C_{\max}\\
\mathrm{s.t.}\quad
&0\leq s_i\leq H,
&&\forall i\in\mathcal{N},\\
&L\leq C_{\max}\leq C^{\mathrm{ub}},\\
&C_{\max}\geq s_i+\tau_{ij}+L,
&&\forall(i,j)\in\mathcal{A},\\
&\sum_{i\in\mathcal{N}}f_i=1,\\
&s_i\leq H(1-f_i),
&&\forall i\in\mathcal{N},\\
&\Xi\in\mathcal{C}_{\mathrm{hd}}\cap
\mathcal{C}_{\mathrm{int}}\cap\mathcal{C}_{\mathrm{mask}},\\
&\mathbf{x}^{e,1}_{ijk}=\widetilde{\mathbf{v}}_{ijk},
&&\forall(i,j,k)\in\mathcal{I},\\
&\mathbf{x}^{e,2}_{ijk}=\mathbf{y}^{e,1}_{ijk},
&&\forall(i,j,k)\in\mathcal{I},\\
&\mathbf{h}_{ijk}=\mathbf{y}^{e,2}_{ijk},
&&\forall(i,j,k)\in\mathcal{I},\\
&\eta_{\nu}=\mathbf{w}_{\nu}\mathbf{x}_{\nu}+b_{\nu},
&&\forall\nu\in\mathcal{V},\\
&\underline{\eta}_{\nu}\leq\eta_{\nu}\leq\overline{\eta}_{\nu},
&&\forall\nu\in\mathcal{V},\\
&y_{\nu}\geq0,\qquad y_{\nu}\geq\eta_{\nu},
&&\forall\nu\in\mathcal{V},\\
&y_{\nu}\leq\overline{\eta}_{\nu}\beta_{\nu},
&&\forall\nu\in\mathcal{V}_{u},\\
&y_{\nu}\leq\eta_{\nu}-\underline{\eta}_{\nu}(1-\beta_{\nu}),
&&\forall\nu\in\mathcal{V}_{u},\\
&y_{\nu}=0,
&&\forall\nu\in\mathcal{V}_{0},\\
&y_{\nu}=\eta_{\nu},
&&\forall\nu\in\mathcal{V}_{+},\\
&r_{ij,r}=\frac{1}{K}
\sum_{k\in\mathcal{K}_{ij}}m_{ijk,r},
&&\forall(i,j)\in\mathcal{A},\\
&&&r=1,\ldots,4,\\
&\mathbf{x}^{\mathrm{head}}_{ij}=
[\kappa_{ij},\mathbf{r}_{ij}^{\mathsf T},q^{\mathrm{sel}}_{ij}/K]^{\mathsf T},
&&\forall(i,j)\in\mathcal{A},\\
&\mathbf{x}^{h}_{ij}=\mathbf{x}^{\mathrm{head}}_{ij},
&&\forall(i,j)\in\mathcal{A},\\
&z_{ij}=(\mathbf{w}^{o})^{\mathsf T}\mathbf{y}^{h}_{ij}+b^{o},
&&\forall(i,j)\in\mathcal{A},\\
&z_{ij}\geq\gamma_{\rho}=\log\frac{\rho}{1-\rho},
&&\forall(i,j)\in\mathcal{A},\\
&f_i,u_{ij},b_{ijk},o_{ijk},a_{ijk}\in\{0,1\},
\\
&\beta_{\nu}\in\{0,1\},
&&\forall\nu\in\mathcal{V}_{u}.
\end{aligned}
\label{eq:app_complete_nnmilp}
\end{equation}

}

\rev{

\section{Detailed Two-Stage ACO Scheduling Procedure}
\label{app:aco_workflow}

This appendix provides the detailed procedure of the two-stage ACO
scheduler introduced in Section~\ref{sec:aco}.
Figure~\ref{fig:aco_workflow} organizes the procedure into three blocks:
surrogate-guided order search, quality-diverse elite selection, and
neural-aware timing refinement with full validation.
The following subsections describe these blocks in detail.

\begin{figure*}[t]
    \centering
    \includegraphics[width=\textwidth]{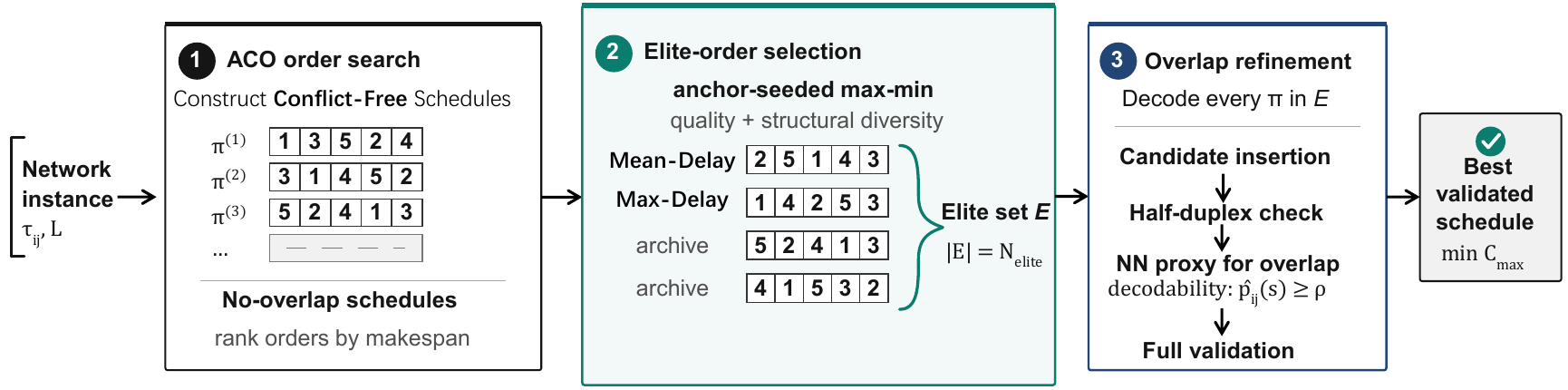}
    \caption{\rev{Workflow of the proposed two-stage ACO heuristic. Block~1 performs
    surrogate order search, Block~2 forms the quality-diverse elite handoff set,
    and Block~3 performs NN-aware controlled-overlap refinement and full
    validation.}}
    \label{fig:aco_workflow}
\end{figure*}

\subsection{Surrogate-Guided Order Search}
Each ant represents a candidate solution as a permutation
\(\boldsymbol{\pi}=(\pi_1,\ldots,\pi_N)\) of the transmitting nodes. Given
an order, the surrogate decoder inserts the nodes sequentially and assigns
each node the earliest nonnegative start time at which all of its transmission
and reception activities occur after the activities induced by the previously
inserted nodes at every physical node. The resulting surrogate schedule is
therefore collision-free and half-duplex feasible by construction, and its
one-round scheduling time provides an inexpensive score for guiding the order
search.

At construction position \(t\), let \(\mathcal{U}_t\) be the set of
unscheduled nodes. For each candidate \(i\in\mathcal{U}_t\), edge pheromone
captures the desirability of placing \(i\) after the previously selected
node, while position pheromone captures the desirability of assigning \(i\)
to position \(t\). These two sources are combined as
\begin{equation}
    \phi_{ti}
    =\left(\tau^{\mathrm{edge}}_{\pi_{t-1},i}\right)^{\omega_e}
     \left(\tau^{\mathrm{pos}}_{ti}\right)^{\omega_p}.
    \label{eq:aco_pheromone}
\end{equation}
Here, \(\phi_{ti}\) is the combined pheromone value,
\(\tau^{\mathrm{edge}}_{\pi_{t-1},i}\) is the edge pheromone associated with
placing node \(i\) after \(\pi_{t-1}\), and \(\tau^{\mathrm{pos}}_{ti}\) is
the position pheromone associated with placing node \(i\) at position \(t\);
their relative contributions are controlled by \(\omega_e\) and \(\omega_p\).
For \(t=1\), a virtual start node replaces \(\pi_{t-1}\).

The propagation-aware heuristic \(\eta_{ti}\) is defined in
\eqref{eq:aco_heuristic}.
Here, \(\Delta C_{ti}\) is the increase in surrogate one-round scheduling time
caused by
selecting node \(i\), \(h_i\) is its propagation tail defined in
Section~\ref{sec:system},
and \(S=L+\max_{p,q\in\mathcal{N}}\tau_{pq}\) is the time-normalization scale.
The parameter \(\lambda\) controls the preference for long propagation tails.
Thus, \(\eta_{ti}\) favors a long-tail node when inserting it causes only a
small increase in the current surrogate one-round scheduling time.

Using \(\phi_{ti}\) and \(\eta_{ti}\), the probability of selecting node
\(i\) at position \(t\) is
\begin{equation}
    P_{ti}=\frac{\phi_{ti}^{\alpha}\eta_{ti}^{\beta}}
    {\displaystyle\sum_{k\in\mathcal{U}_t}
    \phi_{tk}^{\alpha}\eta_{tk}^{\beta}}.
    \label{eq:aco_selection_probability}
\end{equation}
The denominator normalizes the probabilities over all unscheduled nodes,
with \(k\) serving as the summation index. The parameters \(\alpha\) and
\(\beta\) control the influence of the combined pheromone and heuristic,
respectively. A roulette-wheel draw according to \(P_{ti}\) selects the next
node.

After all ants complete their permutations in an iteration, both pheromone
matrices undergo evaporation:
\begin{equation}
    \tau\leftarrow(1-\gamma)\tau,
    \qquad
    \tau\in\{\tau^{\mathrm{edge}},\tau^{\mathrm{pos}}\},
    \label{eq:aco_evaporation}
\end{equation}
where \(\gamma\in(0,1)\) is the evaporation rate. This operation gradually
reduces the influence of previously learned edges and positions, preventing
unbounded pheromone accumulation and allowing the search to adapt as better
orders are discovered. After evaporation, the best orders reinforce their
constituent edges and positions as described next.

Let \(\mathcal{B}=\{\boldsymbol{\pi}^{\mathrm{iter}},
\boldsymbol{\pi}^{\mathrm{global}}\}\) contain the iteration-best and global-best
surrogate orders, respectively. After evaporation, each
\(\boldsymbol{\pi}\in\mathcal{B}\) deposits the amount
\begin{equation}
    Q(\boldsymbol{\pi})
    =\kappa_{\boldsymbol{\pi}}
    \frac{\widetilde{C}(\boldsymbol{\pi}^{\mathrm{global}})}
    {\widetilde{C}(\boldsymbol{\pi})},
    \label{eq:aco_deposit}
\end{equation}
where \(\widetilde{C}(\boldsymbol{\pi})\) is the one-round broadcast scheduling
time returned by the
collision-free surrogate decoder and
\(\kappa_{\boldsymbol{\pi}}>0\) is a reinforcement weight, chosen larger for
the global-best order. The amount \(Q(\boldsymbol{\pi})\) is added to every
edge transition and node--position assignment appearing in that order. This
quality-scaled update gives stronger reinforcement to better orders while
retaining information from the current iteration. After the deposits, all
pheromone values are clipped to fixed bounds
\([\tau_{\min},\tau_{\max}]\) to prevent stagnation and numerical growth.

\subsection{Quality-Diverse Elite Selection}
After the surrogate ACO terminates, all distinct node orders encountered
during the search are retained in an archive and ranked by their surrogate
one-round broadcast scheduling times. The elite set is initialized with two propagation-aware anchor
orders, obtained by sorting the nodes according to their mean propagation
delays and propagation tails, respectively. The remaining elites are selected
from high-ranking archived orders using a max--min diversity criterion that
compares both directed adjacency edges and node positions. This selection
balances surrogate quality with structural diversity, reducing the risk that
the second stage evaluates only a narrow group of similar orders whose proxy
ranking may not remain optimal after NN-aware decoding.

Starting from the two anchor orders, the remaining elites are selected
iteratively from a pool of high-ranking archived orders. Let \(\mathcal{E}\)
be the set of elites already selected. Each candidate order
\(\boldsymbol{\pi}\) is assigned the score
\begin{equation}
    \begin{aligned}
        D(\boldsymbol{\pi}\mid\mathcal{E})
        ={}&\min_{\boldsymbol{\pi}'\in\mathcal{E}}
        d_{\mathrm{edge}}(\boldsymbol{\pi},\boldsymbol{\pi}')\\
        &+\mu_p\min_{\boldsymbol{\pi}'\in\mathcal{E}}
        d_{\mathrm{pos}}(\boldsymbol{\pi},\boldsymbol{\pi}')
        -\mu_r R(\boldsymbol{\pi}).
    \end{aligned}
    \label{eq:elite_diversity_score}
\end{equation}
Here, \(d_{\mathrm{edge}}\) is one minus the Jaccard similarity between the
directed adjacency-edge sets of two orders, including the virtual-start
transition; \(d_{\mathrm{pos}}\) is the
fraction of positions occupied by different nodes, and \(R(\boldsymbol{\pi})\)
is the normalized surrogate rank. The weights \(\mu_p\) and \(\mu_r\) control
the contributions of positional diversity and surrogate quality,
respectively. At each step, the candidate with the largest score in
\eqref{eq:elite_diversity_score} is added to \(\mathcal{E}\) until the required
number of elites is reached.

\subsection{Neural-Aware Timing Refinement and Validation}
Each elite order is decoded by inserting its nodes sequentially. When node
\(\pi_t\) is inserted, the decoder constructs a finite set of candidate start
times between zero and the node's collision-free anchor. In addition to the two
endpoints and a coarse coverage of this interval, the candidate set includes
local samples around propagation-shifted activity times generated by the
previously scheduled nodes. These event-centered samples concentrate the
search near locations where the half-duplex or interference relationships can
change, while the collision-free anchor provides a temporally safe fallback.
The
decoder can therefore explore useful controlled-overlap placements without
densely scanning the entire continuous-time interval.

Each candidate start time is checked incrementally against the current partial
schedule. The decoder verifies the strict half-duplex constraints and
reevaluates only the reception contexts changed by the new transmission,
using the fixed link received-power estimates and the schedule-dependent
arrival offsets under the feature conventions of
Section~\ref{sec:predictor}. Since all
unchanged contexts were validated at earlier insertions, this procedure
preserves partial-schedule feasibility without repeatedly evaluating the
entire schedule.

Among the feasible candidate times, the decoder first selects the one with
the smallest partial one-round scheduling time. Ties are resolved, in order, by the earliest
completion of the inserted node, the largest minimum predicted success
probability, and the earliest start time. This rule keeps one-round scheduling-
time reduction
as the primary objective while using reliability margin only as a tie-breaker.
The collision-free anchor is retained as a fallback candidate, but it must
pass the same NN-based feasibility check; if no feasible insertion exists, the elite
order is discarded.

After an elite order has been fully decoded, the resulting schedule is
validated from scratch by reconstructing all receiver contexts and checking
the complete half-duplex and neural reliability conditions. Among the
validated elite schedules, the one with the smallest \(C_{\max}\) is returned.
If every NN-aware decode fails, the best collision-free surrogate schedule is
considered only after passing the same full validation. Unlike the NN-MILP,
this procedure does not certify global optimality.

}

\fi

\end{document}